\documentclass[modern]{aastex631}
\usepackage{natbib}
\usepackage{graphicx}
\usepackage{mathrsfs}
\usepackage{amsmath}
\usepackage{tikz}
\usepackage{hyperref}
\newcommand{\change}[1]{#1}

\tikzstyle{blue} = [rectangle, rounded corners, minimum width=3cm, minimum height=1cm,text centered, draw=black, fill=blue!30]
\tikzstyle{red} = [rectangle, rounded corners, minimum width=3cm, minimum height=1cm,text centered, draw=black, fill=red!30]
\tikzstyle{green} = [rectangle, rounded corners, minimum width=1.5cm, minimum height=1cm,text centered, draw=black, fill=green!30]
\tikzstyle{orange} = [rectangle, rounded corners, minimum width=1.5cm, minimum height=1cm,text centered, draw=black, fill=orange!30]
\tikzstyle{arrow} = [thick,->,>=stealth]
 
\begin{document}
 
\title{Planets in the Radius Gap Lack the Peas-in-a-Pod Size Correlation}
 
\author[0000-0002-9916-3517]{Quadry Chance}
\affiliation{Department of Astronomy, University of Florida, Gainesville, FL 32611, USA}
\affiliation{NASA Postdoctoral Program Fellow, NASA Goddard Space Flight Center, Greenbelt, MD 20771, USA}
\author[0000-0002-3247-5081]{Sarah Ballard}
\affiliation{Department of Astronomy, University of Florida, Gainesville, FL 32611, USA}
 
\begin{abstract}
Planets in compact multi-transiting systems tend to exhibit self-similarity with their neighbors, a phenomenon commonly called ``peas-in-a-pod.'' Previous studies have identified that this self-similarity appears independently among super-Earths and sub-Neptunes orbiting the same star. In this study, we investigate whether the peas-in-a-pod phenomenon holds for planets in the radius valley between these two categories (located at $\sim 1.8\,R_\oplus$). Employing the \textit{Kepler} sample of planets in multi-transiting systems, we construct a difference-in-differences test that compares the observed fraction of size-similar adjacent pairs to the fraction expected from the underlying radius distribution alone, computed independently for valley-inclusive and valley-exclusive pairs. We find that non-valley pairs exhibit a $1.87\times$ enhancement of size-similar pairs above the baseline, consistent with the well-established peas-in-a-pod phenomenon. Pairs involving a radius valley planet show no such enhancement, and we exclude at $p = 0.001$ the hypothesis that the same size-similarity mechanism operates at the same strength for valley-inclusive pairs. The observed fraction of size-similar valley-inclusive pairs is consistent with independent draws from the radius distribution, with no additional intra-system correlation. We further compare the period ratio distributions for the two pair classes. While globally indistinguishable (KS $p = 0.848$), valley-inclusive pairs cluster near the 3:2 mean-motion resonance at more than twice the rate of the parent population, while avoiding the tightest orbital spacings entirely. The convergence of disrupted size-similarity and anomalous resonance architecture, together with independently measured elevated eccentricities among valley planets, is consistent with a stochastic process such as late-stage giant impacts contributing to the population of planets in the radius valley.
\end{abstract}
 
 
\section{Introduction}
There exists a great diversity of radii among exoplanets, with the most common being a few times the radius of Earth: so-called super-Earths and sub-Neptunes \citep{Borucki11,batalha_2013, Howard12}.
Within the radius distribution of detected exoplanets, the marked decrease in the number of planets with radii $\sim 1.8\,R_\oplus$ is commonly referred to as the
radius gap''\citep{Fulton17, Fulton18}or radius valley.'' We adopt the latter term throughout this work, as it better reflects the observed minimum in the radius distribution without implying a complete absence of planets at this radius.
Our understanding of the valley has grown progressively more detailed: it is also a function of orbital period, host star spectral type, stellar age, and stellar phase space density \citep{Fulton18, berger_revised_2018, berger_gaiakepler_2020, Hardegree19, david_evolution_2021, kruijssen_bridging_2020, ho_shallower_2024}. In addition, its degree of ``emptiness" is a subject of active study, an important consideration given the measurement uncertainty of planetary radii. Some have concluded that the valley is not ``empty" of planets \citep{Fulton18, Lopez2018}, though the degree of emptiness depends sensitively on the precision of the planet radius measurements \citep{vaneylen_asteroseismic_2018, petigura_two_2020, Lopez2018, ho_deep_2023}.  
 
As a defining feature of planet occurrence, the radius valley is a useful fulcrum to investigate formation models (see review by e.g. \citealt{venturini_nature_2020}). Its dependence on orbital period indicates a link between planetary radii and system architecture. Various physical mechanisms have been considered to explain the deficit of planets at $\sim1.8R_{\oplus}$, with different formation and evolution models predicting different relationships. Theories of the provenance of the bimodal radius distribution broadly fall into two categories: either it emerges over time as some atmospheric escape process takes place, or it is a directly imprinted during planet formation (without undergoing major change from atmosphere loss, e.g. \citealt{lee_creating_2022,Lopez2018}). 
 
Within the atmospheric loss framework, the bimodal appearance occurs because planets generally fall into two categories: those that manage to hold on a primordial H/He envelope (sub-Neptunes, approximately $\sim$2.5 $R_{\oplus}$) and those stripped down to (relatively) bare cores (super-Earths, approximately $\sim$1.5 $R_{\oplus}$). In this scenario, all small planets are assumed to accrete a gaseous envelopes of a few percent the mass of the core at formation \citep{Lopez2013}. The proposed atmospheric escape can then take similar forms. Core-powered mass loss, by which the internal luminosity of the planet cooling after its formation is sufficient to induce thermal escape of the atmosphere \citep{Ginzburg2018, gupta_signatures_2020} and photoevaporation, whereby planets are stripped of their atmospheres by XUV flux from the host star \citep{Lopez2013,Owen13,Owen2017,jin_planetary_2014,  vaneylen_asteroseismic_2018} are mechanisms that dominate in different regimes depending on the XUV penetration depth and in some cases occur concurrently \citep{owen24}. The fact that the position of the radius valley depends upon orbital period and stellar mass is a strong indication of an insolation-related phenomenon \citep{VanEylen_2019, Petigura18}. 
\begin{figure*}[ht]
    \centering
    \includegraphics[width=\textwidth]{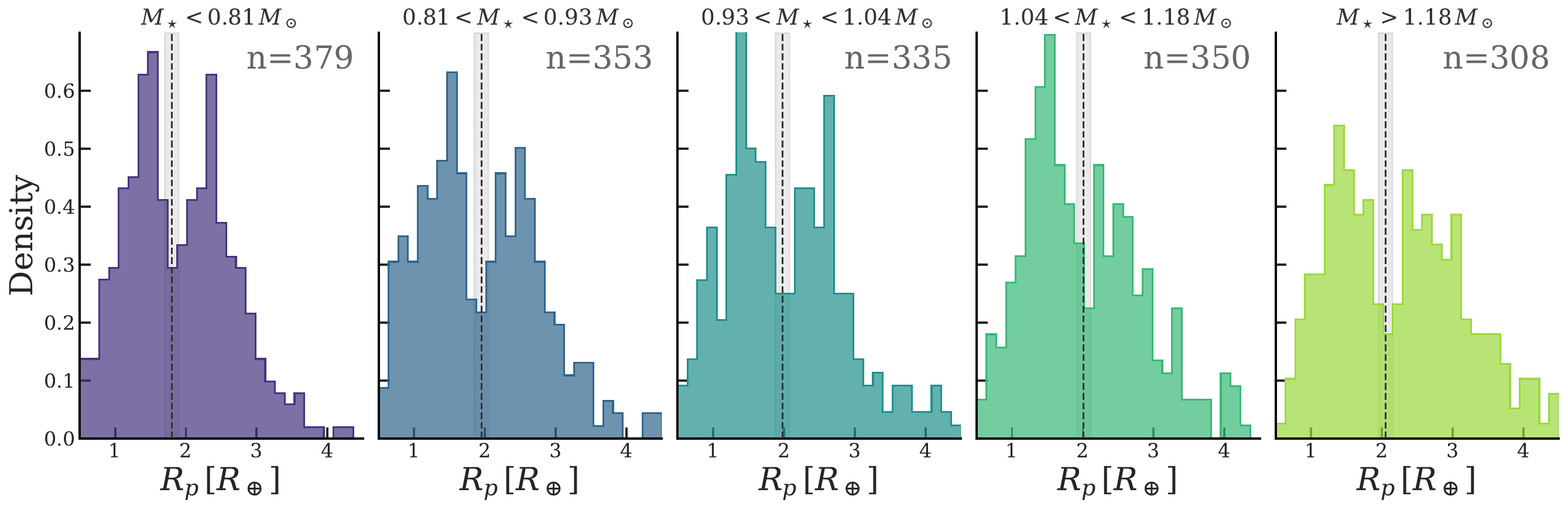}
    \caption{Density distributions of planetary radii in multi-transiting systems, 
    divided into five bins of host star mass (increasing left to right). The dashed 
    line in each panel marks the valley location fitted by \texttt{gapfit} 
    \citep{Loyd20} to that stellar mass bin. The grey shaded band indicates the 
    $\pm 0.1\,R_\oplus$ window used to define valley membership. The number of 
    planets in each bin is noted in the upper right corner of each panel. The valley 
    shifts to larger radii with increasing stellar mass, consistent with previous 
    findings \citep{Fulton18, berger_gaiakepler_2020}.}
    \label{fig:gapfit}
\end{figure*}
 
Other studies have argued that radius valley planets, rather than a population resulting from primordial H/He atmospheric loss, are compositionally distinct or acquired their atmospheres in a different way. One possibility is that radius valley planets are water-rich, at the level of tens of percent \citep{Zeng2019, mousis_irradiated_2020, aguichine_mass-radius_2021, jin_compositional_2018, venturini_nature_2020}. For example, \cite{burn_radius_2024} presented a case for radius valley planets as ``migrated steam worlds" originating outside the snowline and moving inward, a distinct population from the smaller rocky evaporated cores at $\sim1.5R_{\oplus}$. \cite{lee_primordial_2021} demonstrated that atmospheric accretion in the late-stage gas-poor nebula ought to result in a population of planets at a range of radii (including near the valley location). 
 
Another proposed mechanism for atmosphere loss is late-stage giant impacts \citep{Liu15, schlichting_atmospheric_2015, Schlichting18, Biersteker2019}. An impact of a large planetary embryo on a planet with a primordial H/He atmosphere can trigger massive atmosphere loss through both the hydrodynamic ejection of the gas and heating of the planet interior from the collision \citep{Schlichting18}. 
 
\cite{Chance_2022} modeled two atmospheric loss mechanisms for the same set of formation simulations (from \citealt{Dawson16}): photoevaporation and giant impacts. They found that photoevaporation generally results in a nearly empty radius valley, while a non-empty valley points to planets that plausibly originated from giant impacts. Within this framework, valley planets are stripped cores too massive or too far from their host stars to lose their atmospheres to photoevaporation; rather, their atmospheres can be lost only via a giant impact. If this scenario is correct, planets in the radius valley may reflect a dynamically distinct population, whose system architectures bear the imprint of late-stage collisions. In this work, we test this possibility using planet radius and period ratios as demographic probes of disruption.
An important observational constraint on any atmosphere loss mechanism is the slope of the radius valley in period-radius space. \cite{Lopez2018} predicted that giant impacts alone would produce a valley position that increases with orbital period. The observed decrease in valley position with orbital period indicates that giant impacts cannot be the sole origin of the valley, but they may still contribute significantly to its population, particularly for planets too massive or too distant from their host stars to be stripped by photoevaporation.
 
Observationally, the existence of adjacent planets with dramatically different densities \citep{Carter12, Inamdar2016, Bonomo2019} supports the idea that late-stage impacts can lead to divergent outcomes even within the same system. While such dissimilar neighbors do occur, most compact multi-planet systems follow the well-known “peas-in-a-pod” pattern: planets are similar in size, regularly spaced, and often ordered from smaller to larger \citep{Weiss2018, Millholland21,Weiss_2020}.

Planets arising from these disruptive collisions might differ in architectural context from the standard multi-transiting system in ways that have been observed in multiplanet systems already: for example, atypical radius ordering, elevated mutual inclinations or eccentricities,altered spacing, or unusual proximity to resonances \citep{rice_distribution_2024, Gilbert2020, Pu2015a, Millholland2019}.
One emergent consequence of such disruption might, for example, be the period ratio with adjacent planets. If giant impacts disrupt orderly planet formation, we might also expect the well-known period ratio distribution \citep{Fabrycky2014} to appear differently. However, many of these connections are as yet only partly understood. While we might expect a giant impact scenario to produce unusual size ordering, \cite{izidoro_exoplanet_2022} proposed a model by which radius valley planets result from the disruption of resonant chains after the gas disk disperses (see also \citealt{izidoro_formation_2021}). The resulting universal stripping of H/He atmospheres by giant impacts during this instability period ought to produce both the radius valley feature and the peas-in-a-pod phenomenon.
One consequence of such disruption could be a deviation in the period ratio distribution relative to the rest of the population. While we might expect giant impacts to break resonant chains or reverse size-ordering, the specific demographic signature of this scenario remains an open question.
 
Recent work has also shown that planet radius and system dynamics may be connected. In particular, \cite{rice_distribution_2024} demonstrate that the distribution of planet radii in Kepler multi-planet systems depends on the spacing between planets, with more unevenly spaced systems showing more pronounced radius valleys.
Similarly, results from \cite{gilbert_planets_2025} (tentatively) show that planets in the radius valley have systematically higher eccentricities than planets on either side of the valley. They use an adjusted radius to account for the period dependence of the radius valley which further strengthens the robustness of this eccentricity enhancement. This reinforces the idea that the radius valley traces not just a photoevaporative boundary, but possibly a transition in dynamical history.  
 
Much effort has gone into uncovering and modeling the origins of the radius valley. With those foundations in place, we now have an opportunity to study the planets within the valley as a distinct population, in both dynamical and demographic terms.
 
In this work, we investigate whether planets in the radius valley differ from their neighbors in terms of radius and period ratios, and whether they follow the same size-ordering and spacing patterns commonly observed in compact multi-planet systems. This manuscript is organized as follows. In Section~\ref{sec:methods}, we describe our sample construction and the criteria for identifying valley planets. In Section~\ref{sec:analysis}, we present the radius ratio distributions (Section~\ref{sec:rr_distribution}), separate the intrinsic and physical contributions to size-similarity (Sections~\ref{sec:geometric_physical}--\ref{sec:significance}), and compare the period ratio distributions for valley-inclusive and non-valley pairs (Section~\ref{sec:period_ratios}). In Section~\ref{sec:discussion}, we consider physical interpretations, before concluding in Section~\ref{sec:conclusions}.
 
\section{Methods}
\label{sec:methods}

\subsection{Sample selection}

\change{We uniformly draw our planet and host star parameters from Berger et al. (2023). These properties were homogeneously derived using isochrones and Gaia Data Release 3 photometry (Andrae et al. 2023; Fouesneau et al. 2023), Gaia Data Release 3 parallaxes (Lindegren et al. 2021; Vallenari et al. 2023), and spectrophotometric metallicities whenever they were available. Given that we are considering the radius and period ratios of neighboring planets, we select only multi-transit systems from Berger et al. (2023): this sample comprises a total of 1719 planets orbiting 690 host stars.}

\begin{figure*}
\centering
\includegraphics[width=\textwidth]{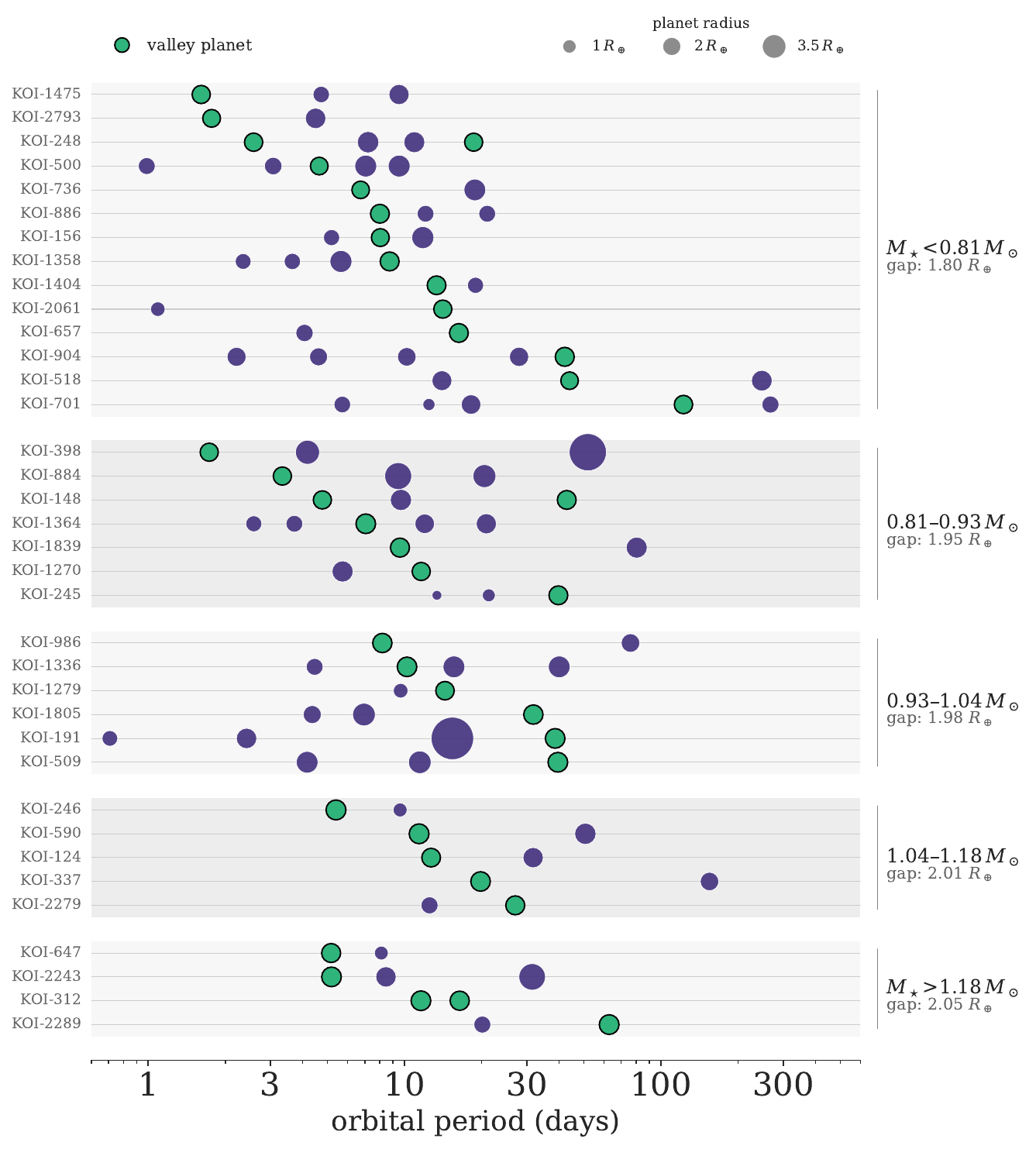}
\change{\caption{The sample of 36 host systems containing at least one radius valley
planet, organized by stellar mass bin (right margin) following
\citet{berger_gaiakepler_2020}.  Radius valley planets (green)
satisfy our membership criteria (Section~\ref{sec:valley membership}):
$|R_p - R_{vc}| < 0.1\,R_\oplus$ for the gap center $R_{vc}$ of the
system's stellar mass bin (gap location indicated at right), and
fractional radius uncertainty $\leq 5\%$. All other transiting planets
in these systems (purple) form the parent distribution against which we
test for size and period-ratio differences in Section~\ref{sec:analysis}.
Within each mass bin, systems are ordered by the period of their innermost
valley planet. Three systems (KOI-148, KOI-248, KOI-312) host two valley
planets each.}
\label{fig:valley_family}}

\end{figure*}
\subsection{Establishing valley membership}
\label{sec:valley membership}
 
We aim to determine whether there exists a difference in the distribution of adjacent planet radius ratios between two samples: systems containing a ``radius valley planet'' and systems without one. This experiment requires identifying a sample of planets in the ``radius valley.'' This is necessarily an exercise with some uncertainty, as the location of the radius valley, somewhere between 1.5-2.0 $R_{\oplus}$ \citep{Fulton17, Hsu19} is a subject of active study. It shifts as a function of stellar mass \citep{Fulton18, berger_gaiakepler_2020} in ways that also covary with insolation, age, and potentially stellar metallicity \citep{Petigura22, ho_deep_2023}. 
 
\begin{table*}[t]
    \centering
    \begin{tabular}{c|c|c|c}
    Stellar mass     & B20 gap location [$R_{\oplus}$]& Gap location, this work [$R_{\oplus}$]\\    
         \hline
    $M_{\star}<0.81M_{\odot}$     & 1.67$\pm$0.07 & 1.80$\pm$0.10\\
     $0.81M_{\odot}<M_{\star}<0.93M_{\odot}$  & 1.87$\pm$0.05 & 1.95$\pm$0.10\\  
     $0.93M_{\odot}<M_{\star}<1.04M_{\odot}$ &  1.89$\pm$0.05& 1.98$\pm$0.10\\
      $1.04M_{\odot}<M_{\star}<1.18M_{\odot}$ & 1.87$\pm$0.06& 2.01$\pm$0.10\\
      $M_{\star}>1.18M_{\odot}$ & 2.05$\pm$0.07 & 2.05$\pm$0.10 \\
    \end{tabular}
    \caption{Location of the radius valley identified with the package \texttt{gapfit}, for multi-transiting systems. We employ the same five stellar mass bins as \cite{berger_gaiakepler_2020}.}
    \label{tbl:gap_location}
\end{table*}
 
In addition to apparent variability in the location of the valley, the estimates of planetary radii themselves vary depending on the stellar parameters. For example, the same \textit{Kepler} planets are on average $\sim$0.05$R_{\oplus}$ larger in the \cite{berger_gaiakepler_2020} catalog than in \cite{berger_gaia-kepler-tess-host_2023}, with the discrepancy growing to $\sim 0.1\,R_{\oplus}$ for stars with $M_{\star}>1.18\,M_{\odot}$. In this sense, whether a planet appears to reside in the radius valley will vary for both astrophysical reasons (e.g.\ its dependence on stellar mass) and non-astrophysical reasons (e.g.\ the provenance of the stellar parameters used to characterize the planets). For this reason, we employed the package \texttt{gapfit} \citep{Loyd20} to fit the location of the valley for our exact planetary samples. Following \cite{berger_gaiakepler_2020}, we form five bins of stellar host mass and fit the valley location to each: one sample for stars $M_{\star}<0.81\,M_{\odot}$, one sample for stars $0.81<M_{\star}<0.93\,M_{\odot}$, one sample for $0.93\,M_{\odot}<M_{\star}<1.04\,M_{\odot}$, one sample for $1.04\,M_{\odot}<M_{\star}<1.18\,M_{\odot}$, and finally one sample for $M_{\star}>1.18\,M_{\odot}$. By employing the resulting valley location for each subsample, we can be sure that our criterion for residing in the valley is approximately correct for that exact sample of planets. Figure~\ref{fig:gapfit} shows the distribution of planetary radii orbiting stars in each mass bin. 
 
By virtue of considering the radius ratio of neighboring planets, we are concerned with the location of the valley for multi-planet systems. We observed it to be slightly offset from the location of the valley reported with the \cite{berger_gaiakepler_2020} sample, by $\sim 0.05$--$0.10\,R_{\oplus}$ (in all stellar mass bins except for the most massive). This might be attributable to the change in planetary parameters between \cite{berger_gaiakepler_2020} and \cite{berger_gaia-kepler-tess-host_2023}, but we find that multiplicity might play a role as well. In breaking the sample into only the multi-transit systems (considered in this paper) versus the singly-transiting systems, we find that the locations of the valley from \cite{berger_gaiakepler_2020} furnish a good fit to the singly-transiting systems (and these are the majority of the sample). It is only for the multi-transiting systems that we identify the valley to be located at a slightly larger planet size. A consideration of this phenomenon is outside the scope of the present manuscript: simple identification of the valley location within our sample of planets is sufficient for our experiment, but we note it as a point of potential future interest. 
 
We employ the following criteria for valley planet membership for this analysis: (1) the planet must reside in a multi-transit system, (2) the $1\sigma$ confidence interval for $R_{p}$ overlaps with the region within $0.1\,R_{\oplus}$ from the location of the valley (using the location corresponding to the host star's mass), and (3) the fractional radius uncertainty is $\leq 5\%$. We craft these criteria to trade off between multiple considerations: while a less stringent error requirement on the planetary radius would increase the sample size, it would also potentially dilute any signal presented by valley planets, as the relative confidence of valley membership decreases with increasing radius uncertainty. We investigate the effects of relaxing this assumption in Section~\ref{sec:relaxation}.
 
Using these three criteria, we identify 39 planets orbiting 36 host stars, including three systems (KOI-312, KOI-148, KOI-248) in which two planets independently satisfy the valley membership criteria. Among these three systems, only KOI-312 has them in adjacent orbits, contributing a valley-valley pair; in the other two the valley planets are non-adjacent. All such systems contribute mixed pairs under this scheme, as each pair still involves a valley member adjacent to another planet. This classification yields 769 non-valley pairs and 39 mixed pairs. The orbital architectures of these 36 systems are shown in Figure~\ref{fig:valley_family}; valley planets span a wide range of orbital periods and often differ noticeably in size from their immediate neighbors, motivating the analysis that follows. They are drawn mostly from the bin corresponding to the least massive stars: 17 of the 39 orbit stars with $M_{\star} < 0.81\,M_{\odot}$, with between 4-7 valley planets per bin for the higher stellar masses shown in Table~\ref{tbl:gap_location}. Our ``parent'' distribution, which comprises our control sample for this experiment, includes all planets in multi-transit systems that do not meet criteria (2) or (3).
 
We construct pairs of adjacent planets, defined as planets that are neighbors in orbital period with no intervening transiting planet in the same system. We classify each adjacent pair as either ``non-valley,'' in which neither planet resides in the radius valley, or ``mixed,'' in which at least one planet is a valley planet. The K00312 system, which contains two valley planets, contributes mixed pairs under this scheme, as each pair still involves a valley member adjacent to another planet. This classification yields 769 non-valley pairs and 39 mixed pairs.

\begin{figure*}[ht]
    \centering
    \includegraphics[width=\textwidth]{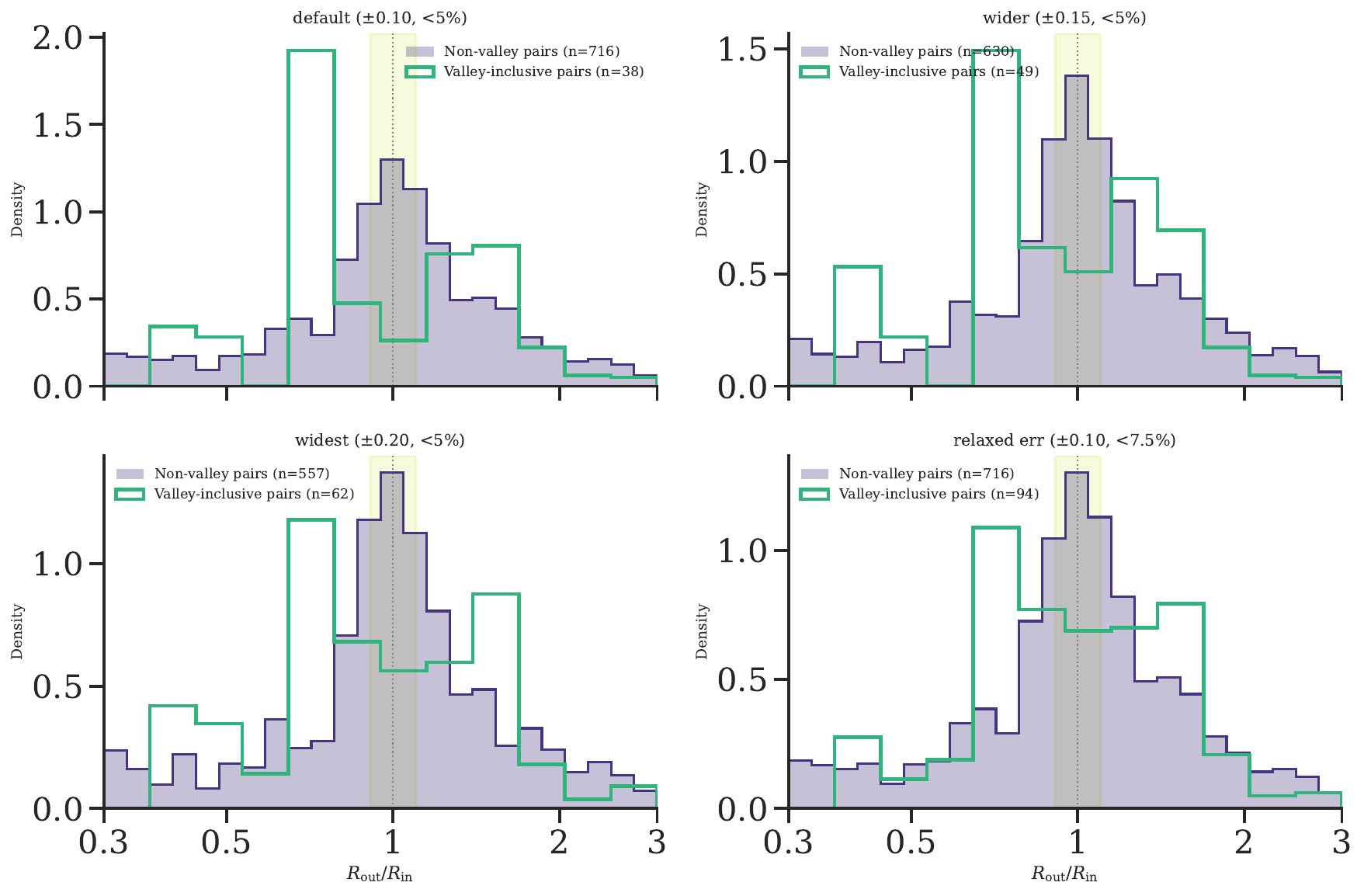}
    \caption{Radius ratio distributions $R_{\rm out}/R_{\rm in}$ for non-valley pairs 
    (non-valley; purple) and valley-inclusive pairs (mixed; green) under progressively relaxed valley membership criteria. Top left: the default window of $\pm0.10\,R_\oplus$ with fractional radius uncertainty $<5\%$ ($n = 716$, $38$). Top right: widened to $\pm0.15\,R_\oplus$ ($n = 630$, $49$). Bottom left: widened further to $\pm0.20\,R_\oplus$ ($n = 557$, $62$). Bottom right: the default spatial window with the uncertainty threshold relaxed to $<7.5\%$ ($n = 716$, $94$). The $x$-axis is logarithmic; the yellow band marks the size-similar region 
    $|\log(R_{\rm out}/R_{\rm in})| < \log(1.1)$, and the dotted vertical line marks 
    exact size equality. The non-valley distribution is shown as a filled histogram; the 
    mixed distribution as a step histogram with coarser bins appropriate for the smaller sample. In all cases the non-valley distribution peaks near unity while the mixed distribution shows elevated density at $R_{\rm out}/R_{\rm in} \sim 0.7$. The deficit near unity in mixed pairs is most pronounced under the most stringent criteria and diminishes with looser classification, consistent with dilution by ambiguous valley members.}
    \label{fig:relaxation_tests}
\end{figure*}
 
\subsection{Effects of changes to valley criteria}
\label{sec:relaxation}
 
We consider the effects of relaxing the criteria for valley membership. Widening the valley from $\pm 0.1\,R_\oplus$ to $\pm 0.15\,R_\oplus$ and then $\pm 0.20\,R_\oplus$ to either side of the fitted valley center increases the sample size by factors of approximately 1.5 and 2, respectively, but the deficit at $R_\mathrm{outer}/R_\mathrm{inner} = 1.0$ becomes less pronounced in each case. Similarly, relaxing the fractional radius uncertainty requirement from $\leq 5\%$ to $\leq 7.5\%$ roughly doubles the number of valley planets but smooths the deficit, consistent with dilution from planets whose valley membership is less secure. In all cases, the signal is strongest among pairs adjacent to the valley planet itself: when we include non-adjacent pairs from systems containing a valley planet, the peak near unity returns, indicating that the departure from self-similarity is localized to the immediate neighborhood of the valley planet. We also find that the deficit persists regardless of whether the valley planet is the innermost or outermost member of the pair, though we defer a more detailed consideration of size-ordering asymmetry to Section~\ref{sec:discussion}. These experiments indicate that the architectural signal is concentrated among the most confidently identified valley planets and their immediate neighbors. We show the radius ratio distributions under each relaxation in Figure~\ref{fig:relaxation_tests}. The quantitative impact on our main result is presented in Section~\ref{sec:robustness}.
 
\subsection{Quantifying size-similarity}
\label{sec:peas_method}
 
To assess whether adjacent planets exhibit size-similarity beyond what the underlying 
radius distribution would predict, we construct a test that separates the geometric 
expectation from any physical peas-in-a-pod enhancement.

For a given class of planet pairs, we define a pair as a "peas" pair if 
$|\log(R_\mathrm{outer}/R_\mathrm{inner})| < \log(1.1)$, corresponding to planets 
that differ in radius by less than 10\%. We then compute two quantities: the observed 
fraction of peas pairs in the data, $f_\mathrm{obs}$, and the expected fraction under 
a geometric null model, $f_\mathrm{null}$.

In the null model, we test whether adjacent planet radii are correlated
beyond what the underlying radius distribution predicts. The procedure
differs by pair class:
\change{\begin{itemize}
\item \textit{Non-valley pairs:} Both radii are drawn independently and
  at random from the pool of all non-valley planets. Neither draw is
  conditioned on any system membership.
\item \textit{Valley-inclusive pairs:} One radius is drawn at random from
  the valley planet pool (representing the valley planet in the pair),
  and the second radius is drawn independently from the \emph{full}
  planet distribution (representing the adjacent non-valley planet, which
  may itself have any radius). Planets excluded from the valley sample
  under criteria (2) or (3) are therefore still eligible to be drawn as
  the non-valley partner in this null model.
\end{itemize}
This construction ensures that $f_{\rm null}$ for each pair class accounts
for the observed rarity of valley-sized planets without imposing any
additional intra-system correlation. We estimate $f_{\rm null}$ via
bootstrap resampling with $N = 5000$ draws.}

The peas-in-a-pod enhancement is then
\begin{equation}
\Delta = f_\mathrm{obs} - f_\mathrm{null}.
\end{equation}
This quantity is zero if adjacent planet radii are uncorrelated, and positive if 
adjacent planets are more similar in size than expected from independent draws. Because 
the null model draws from population-specific radius pools, the baseline for each pair 
class already reflects the shape of the relevant radius distribution—including the 
rarity of planets near the valley.

We complement this with a "widget" test that asks whether the enhancement observed in 
one pair class is consistent with the enhancement in another. Specifically, we compute 
the predicted $f_\mathrm{peas}$ for mixed pairs under two models: a multiplicative 
model, in which the peas-in-a-pod mechanism amplifies $f_\mathrm{null}$ by the same 
factor observed for non-valley pairs, and an additive model, in which the same absolute 
$\Delta$ is applied. Comparing the predicted $f_\mathrm{peas}$ to the observed value 
allows us to test whether the same size-similarity mechanism operates across pair 
classifications.
 
\section{Analysis}
\label{sec:analysis}
 
In this Section, we examine the radius and period ratio distributions corresponding to our sample of valley planets, in comparison to those of the non-valley parent population. In Section~\ref{sec:rr_distribution}, we compare the radius ratio distributions for mixed and non-valley pairs. We investigate the effect of relaxing our valley membership criteria in Section~\ref{sec:relaxation}. In Section~\ref{sec:geometric_physical}, we apply the difference-in-differences test defined in Section~\ref{sec:peas_method} to separate observational and physical contributions to the radius ratio distributions. In Section~\ref{sec:significance}, we report the statistical significance of the disruption to size-similarity among valley planets. In Section~\ref{sec:period_ratios}, we compare the period ratio distributions for mixed and non-valley pairs.

\subsection{Radius ratio distribution}
\label{sec:rr_distribution}
 
Following \cite{Millholland17} and \cite{Weiss2018}, we calculate the radius ratio as the radius of the outer planet to the inner planet ($R_\mathrm{outer}/R_\mathrm{inner}$) for each adjacent pair. As described in Section~\ref{sec:valley membership}, our sample comprises 769 non-valley pairs and 39 mixed pairs.
 
We show the radius ratio distributions for both pair classes in Figure~\ref{fig:spaghetti_null}. The non-valley distribution peaks strongly at $R_\mathrm{outer}/R_\mathrm{inner} = 1.0$, consistent with the well-established peas-in-a-pod phenomenon \citep{Millholland17, Weiss2018, He19,Zhu2021}. A tail toward increasing radius ratios reflects the typical pattern in which outer planets tend to be modestly larger than their inner neighbors. The mixed distribution, by contrast, exhibits a deficit near unity, with peaks at $R_\mathrm{outer}/R_\mathrm{inner} \sim 0.7$--$0.8$ and $\sim 1.3$. These features suggest that valley planets tend to differ substantially in size from their immediate neighbors rather than resembling them.

\begin{table}[h]
\centering
\caption{Size-similarity enhancement for adjacent planet pairs}
\label{tab:main_result}
\begin{tabular}{lccccccc}
\hline
Pair class & $n$ & Observed rate & Null rate & \change{Null 95\% CI} & Enhancement & $p$-value \\
\hline
Non-valley       & 769 & 0.212 & 0.114 & \change{$[0.091, 0.137]$ }& $+0.098$ & $<0.001$ \\
Valley-inclusive &  39 & 0.077 & 0.129 & \change{$[0.026, 0.231]$} & $-0.052$ & 0.899    \\
Contrast         & --- & ---   & ---   & ---              & $+0.151$ & 0.0002   \\
\hline
\end{tabular}
\change{\tablecomments{The wide 95\% CI on the valley-inclusive null $[0.026, 0.231]$
reflects shot noise from the small valley planet pool ($n = 39$); the null
rate is uncertain at this sample size, and the valley-inclusive
result lacks statistical power when considered in isolation ($p = 0.899$).
The overall significance ($p = 0.0002$) derives from the
difference-in-differences contrast with the precisely constrained
non-valley null $[0.091, 0.137]$.}}
\end{table}
 
We note that this raw comparison does not account for the fact that the expected radius ratio distribution depends on the underlying planet radius distribution, and in particular on the rarity of planets near $\sim 1.8\,R_\oplus$. A pair involving a valley planet is intrinsically less likely to exhibit a radius ratio near unity even in the absence of any physical disruption of size-similarity. In the following subsections, we disentangle this contribution from the physical peas-in-a-pod enhancement.

\subsection{Separating intrinsic and physical contributions to the radius ratio distribution}
\label{sec:geometric_physical}
 
The raw radius ratio distributions presented in Section~\ref{sec:rr_distribution} 
do not account for the fact that the expected distribution depends on the underlying 
planet radius distribution. In particular, because valley planets occupy a narrow and 
underpopulated region near $\sim 1.8\,R_\oplus$, pairs involving a valley planet are 
intrinsically less likely to exhibit radius ratios near unity even in the absence of 
any disruption to size-similarity. To isolate the physical peas-in-a-pod enhancement 
from this base-rate effect, we apply the test described in Section~\ref{sec:peas_method} 
to both the non-valley and mixed pair classes independently, computing a 
population-specific null $f_\mathrm{null}$ for each.

\begin{figure*}[ht]
    \centering
    \includegraphics[width=\textwidth]{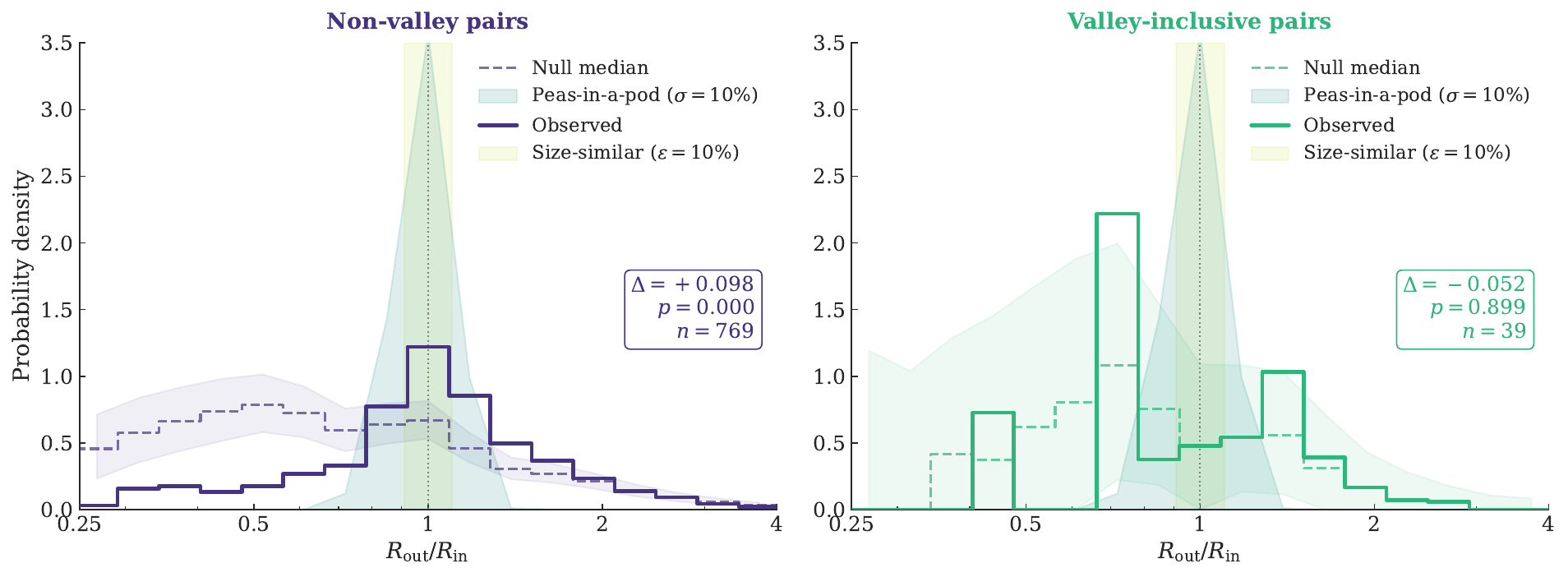}
    \caption{Observed radius ratio distributions (solid histograms) compared to the 
    null distribution for non-valley pairs (left; $n = 769$) and valley-inclusive pairs (right; $n = 39$), plotted on a logarithmic axis. The dashed histogram shows the median of the shuffle null (planet radii drawn independently from the respective population radius pool, $N = 5000$ bootstrap draws), with the shaded band indicating the 95\% bootstrap interval. The dotted histogram shows the expected distribution if a perfect peas-in-a-pod mechanism operated at its characteristic width ($\sigma = 10\%$). The yellow band marks the size-similar region ($\varepsilon = 10\%$). Non-valley pairs show a significant excess of size-similar pairs above the shuffled pairs null ($\Delta = +0.098$, $p < 0.001$), while valley-inclusive pairs fall within their null ($\Delta = -0.052$, $p = 0.899$), exhibiting no peas-in-a-pod enhancement.}
    \label{fig:spaghetti_null}
\end{figure*}

Figure~\ref{fig:spaghetti_null} shows the result. For non-valley pairs (left panel), the observed distribution (solid histogram) 
exceeds the shuffle null (dashed) throughout the size-similar region, confirming 
a strong physical enhancement of radius ratios near unity beyond what independent 
draws from the non-valley radius pool would produce. For mixed pairs (right panel), 
the observed distribution is statistically indistinguishable from the shuffle null 
constructed from the valley and non-valley radius pools. The rarity of valley-sized 
planets alone predicts a low rate of size-similar pairs, and the data show no 
enhancement above this already-low baseline. In contrast, non-valley pairs exceed 
their baseline by a factor of $1.86\times$. The question is therefore not why 
valley-inclusive pairs have fewer size-similar pairs than non-valley pairs in 
absolute terms (the observed planet radius distribution guarantees this) but why the 
peas-in-a-pod mechanism that boosts non-valley pairs above their baseline is 
entirely absent for valley-inclusive pairs. We quantify the significance of this 
contrast in Section~\ref{sec:significance}.
 
\begin{figure}[ht]
    \centering
    \includegraphics[width=\columnwidth]{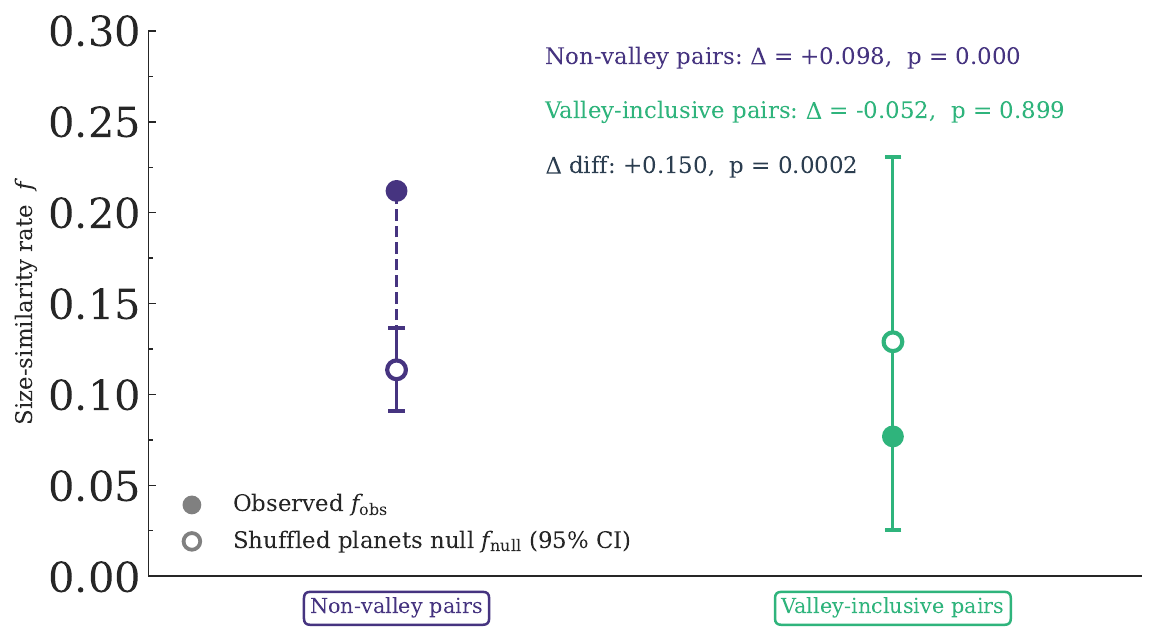}
    \caption{The peas-in-a-pod ``widget'' for non-valley and valley-inclusive pairs. 
    Filled circles show the observed size-similarity rate $f_\mathrm{obs}$; open 
    circles show the null rate $f_\mathrm{null}$ from shuffled planet radii drawn 
    independently from each population's radius pool ($N = 5000$ bootstrap draws), 
    with 95\% confidence intervals. The dashed line connecting each pair of circles 
    visualizes the enhancement $\Delta = f_\mathrm{obs} - f_\mathrm{null}$. 
    Non-valley pairs show a significant excess of size-similar pairs above the shuffled 
    pairs null ($\Delta = +0.098$, $p < 0.001$), while valley-inclusive pairs fall below their 
        null ($\Delta = -0.052$, $p = 0.899$), exhibiting no peas-in-a-pod enhancement.}
    \label{fig:size_similarity}
\end{figure}
\subsection{Statistical significance of disrupted self-similarity}
\label{sec:significance}
The previous section showed qualitatively that valley-inclusive pairs lack the 
excess near $R_\mathrm{outer}/R_\mathrm{inner} = 1$ seen among non-valley pairs. 
We now quantify this contrast. The well-established peas-in-a-pod phenomenon 
implies the existence of some mechanism (a ``widget'', if you will) that makes adjacent 
planets more similar in size than independent draws from the radius distribution 
would predict. The strength of this widget is measured by the enhancement 
$\Delta = f_\mathrm{obs} - f_\mathrm{null}$: positive if the widget is active, 
zero if adjacent planet radii are uncorrelated. Crucially, the null rate 
$f_\mathrm{null}$ is computed separately for each pair class using 
population-specific radius pools, so the rarity of valley-sized planets is already 
accounted for in the baseline. An enhancement above this baseline reflects the 
action of the widget itself, not the shape of the radius distribution. Under the 
null hypothesis, the widget operates uniformly across all planet pairs, and 
valley-inclusive pairs should exceed their baseline by the same factor as 
non-valley pairs. Table~\ref{tab:widget_test} reports the observed and null 
size-similarity rates for each pair class.
\begin{figure}[ht]
    \centering
    \includegraphics[width=\columnwidth]{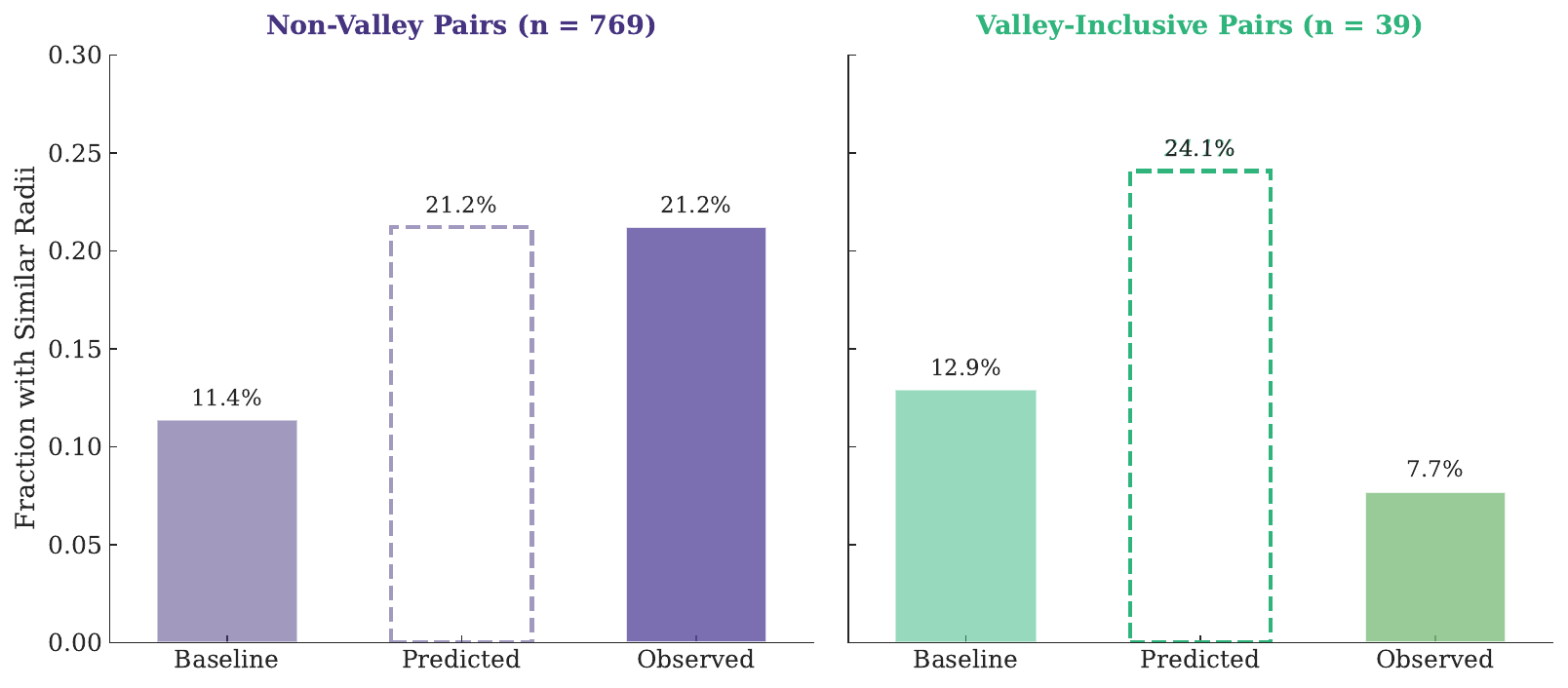}
    \caption{The ``widget'' test comparing the baseline, predicted (if 
    the peas-in-a-pod widget operated at the same strength as in non-valley pairs), and observed 
    size-similarity fractions for non-valley pairs (left; $n = 769$) and valley-inclusive pairs (right; $n = 39$). For non-valley pairs the observed fraction ($21.2\%$) matches the predicted value exactly, confirming the peas-in-a-pod enhancement. For valley-inclusive pairs the baseline is $12.9\%$ and the predicted fraction under the same enhancement would be $24.1\%$, but the observed fraction is only $7.7\%$, falling below even the baseline. This excludes the hypothesis 
    that the same size-similarity mechanism operates at the same strength for 
    valley-inclusive pairs.}
    \label{fig:widget_test}
\end{figure}
\begin{table}[h]
\centering
\caption{Does the peas-in-a-pod mechanism operate at the same
strength for valley-inclusive pairs? \change{The logit model applies the enhancement
as a multiplicative odds ratio in log-odds space, keeping the predicted rate
bounded in $[0,1]$. It is the most conservative of the three models; all
three nonetheless reject the hypothesis that the same mechanism operates at
the same strength for valley-inclusive pairs.}}
\label{tab:widget_test}
\begin{tabular}{lccc}
\hline
Model & Predicted rate & Observed rate & $p$-value \\
\hline
Multiplicative ($\times1.87$) & 0.241 & 0.077 & $<0.001$ \\
Additive ($+0.098$)           & 0.227 & 0.077 & $<0.001$ \\
\change{Logit/odds ratio ($\times2.10$ in odds space) & 0.237 & 0.077 & 0.009} \\
\hline
\end{tabular}
\end{table}
 
Among non-valley pairs ($n = 769$), we find $f_\mathrm{obs} = 0.212$ and 
$f_\mathrm{null} = 0.114$, yielding an enhancement of $\Delta = +0.098$ 
($p < 0.0001$). This corresponds to a factor of $1.86\times$ more similar pairs 
than expected from independent draws, consistent with the well-established 
peas-in-a-pod phenomenon \citep{Millholland17, Weiss2018, Zhu2021}. Among 
valley-inclusive pairs ($n = 39$), we find $f_\mathrm{obs} = 0.077$ and 
$f_\mathrm{null} = 0.129$, yielding $\Delta = -0.052$ ($p = 0.899$). Even 
after setting the bar low to account for the scarcity of valley-sized planets, 
valley-inclusive pairs do not clear it. The difference in enhancement between 
the two classes, $\Delta_\mathrm{non\text{-}non} - \Delta_\mathrm{mixed} = 
+0.151$, is significant at $p = 0.0002$. 
Figure~\ref{fig:size_similarity} summarizes this comparison.
 
We further test whether the valley-inclusive result is consistent with the same 
widget operating at the strength observed for non-valley pairs 
(Table~\ref{tab:widget_test}; Figure~\ref{fig:widget_test}). Under a 
multiplicative model, in which the widget amplifies the baseline by the same 
factor of $\times 1.87$ measured for non-valley pairs, we would predict 
$f_\mathrm{peas} = 0.241$, compared to the observed $0.077$. We reject this 
hypothesis at $p = 0.001$. Under an additive model, in which the same absolute 
enhancement $\Delta = +0.098$ is applied to the valley-inclusive baseline, the 
predicted $f_\mathrm{peas} = 0.227$ is rejected at $p = 0.001$. It is not 
merely that we fail to detect the widget among valley-inclusive pairs; we can 
positively exclude that it operates at the same strength as in the broader 
population. Whatever mechanism produces intra-system size similarity among super-Earths and sub-Neptunes appears not to operate, or operates at significantly reduced strength, for pairs involving a valley planet.
\change{We would be remiss not to note that the valley-inclusive sample is modest in
size (n = 39 pairs). Considered in isolation, the valley-inclusive result
$(\Delta = -0.052, p = 0.899)$ is not itself statistically significant: the
observed fraction falls within the 95\% bootstrap interval of the null
distribution. The overall significance of the difference-in-differences
contrast (p = 0.0002) derives not from an improbably low valley-inclusive
rate in isolation, but from the contrast between a strong enhancement among
non-valley pairs and a null result among valley-inclusive pairs. This framing
is important: the result is a statement about the \emph{difference} in the
operation of the peas-in-a-pod mechanism between the two pair classes, not
a claim that the valley-inclusive rate is intrinsically extreme. The small
sample size makes the valley-inclusive result fragile: the observed count of
3 size-similar pairs out of 39 (or 38 in the bootstrap test; see
Section~\ref{sec:robustness}) corresponds to $f_{\rm obs} = 0.077$.
Observing one additional size-similar valley pair would bring the rate to
$\approx 0.098$ (4/41), matching the null expectation, and two additional
pairs would approach the null. Roughly five additional pairs above the
current count (8/39 $\approx 0.205$) would approach the enhancement level
observed in non-valley systems. While the statistical contrast remains
significant (p = 0.0002), the small sample underscores the need for future
studies with expanded samples to confirm whether the disrupted size-similarity
among valley planets is a robust feature of this population. A forward-looking
robustness assessment is provided in Section~\ref{sec:robustness}.}

\subsection{Period ratio distributions of radius valley and non-valley planets}
\label{sec:period_ratios}
 
We turn now to the distribution of period ratios between adjacent planets. Because orbital periods are precisely measured, this observable is free of the intrinsic complications affecting the radius ratio analysis: the period ratio distribution has no dependence on the underlying planet radius distribution. We compute $P_\mathrm{outer}/P_\mathrm{inner}$ for each adjacent pair and compare the mixed and non-valley samples directly.
\begin{figure*}[ht]
    \centering
    \includegraphics[width=\textwidth]{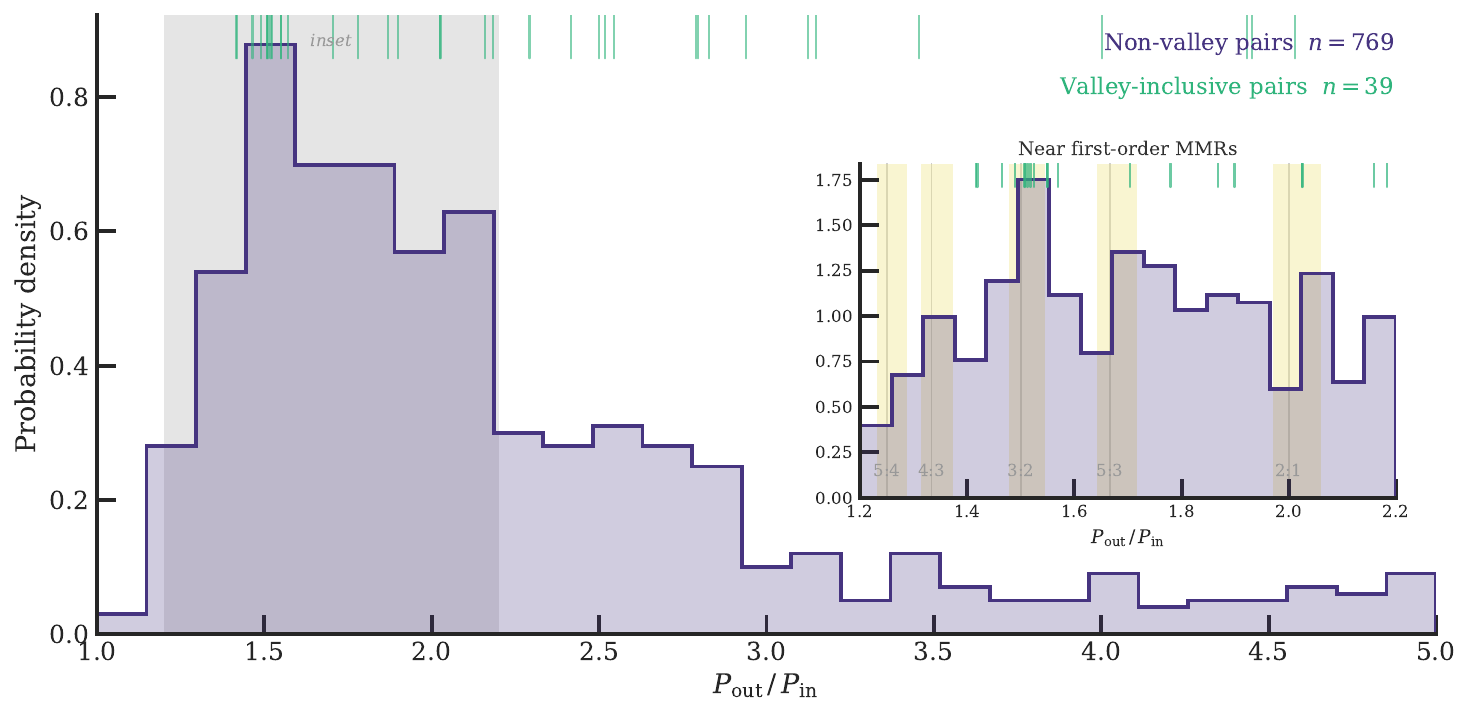}
    \caption{Period ratio distributions $P_{\rm out}/P_{\rm in}$ for adjacent non-valley pairs (purple; $n = 769$) and valley-inclusive pairs (green tick marks at top; $n = 39$). The shaded grey region indicates the range shown in the inset. 
    \textit{Inset:} Zoom into the region near first-order mean-motion resonances 
    ($P_{\rm out}/P_{\rm in} = 1.2$--$2.2$). Vertical grey lines mark exact 
    commensurabilities; yellow bands show the asymmetric near-resonance window of 
    \citet{dai_prevalence_2024} ($-0.015 \leq \Delta \leq 0.03$). Valley-inclusive pairs cluster 
    near the 3:2 resonance at more than twice the rate of non-valley pairs 
    (6/39, 15\% vs.\ 50/769, 7\%; $p = 0.09$). A two-sample 
    Kolmogorov-Smirnov test on the full distributions yields $p = 0.848$, confirming 
    that the global period ratio distributions are statistically indistinguishable; 
    the resonance excess is a localized feature.}
    \label{fig:period_ratios}
\end{figure*}
A two-sample Kolmogorov-Smirnov test yields $p = 0.848$, indicating that the global period ratio distributions for mixed and non-valley pairs are statistically indistinguishable. Radius valley planets are not drawn from a fundamentally different region of period ratio space. Despite this global similarity, however, the local structure of the groups differs near resonance.
 
We assess proximity to mean-motion resonance using the criterion of \cite{dai_prevalence_2024}, who define a pair as near-resonant if its period ratio lies within the asymmetric boundary $-0.015 \leq \Delta \leq 0.03$ of exact commensurability. Mixed pairs show clustering near the 3:2 resonance: $15.4\%$ of mixed pairs ($6/39$) fall within this window, compared to $6.5\%$ of non-valley pairs 
($50/769$; Fisher exact two-sided $p = 0.09$). We report a two-sided test 
here, as we had no prior expectation of excess at a specific resonance. At 
tighter resonances (5:4, 4:3, and 5:3 combined), only $2.6\%$ of mixed pairs 
($1/39$) are found, compared to $10.5\%$ of non-valley pairs ($81/769$; 
one-sided $p = 0.078$). A one-sided test is appropriate here, as the giant 
impact framework predicts that dynamically perturbed systems should avoid 
close orbital spacings. While the latter result is marginal, the direction is consistent with a population that avoids the tightest orbital spacings. No mixed pairs have period ratios below the dynamical packing criterion of $P_\mathrm{outer}/P_\mathrm{inner} = 1.33$ \citep{wu_dynamical_2019}, compared to $5\%$ of non-valley pairs ($39/769$). The absence of any mixed pairs below this threshold, out of 39 total, is striking on its own.
 
Taken together, these results suggest that radius valley planets cluster near the widest first-order resonance (3:2) while avoiding the tightest orbital configurations entirely. This pattern is consistent with a dynamically perturbed population that has relaxed toward the most stable wide resonance. The contrast between global similarity (KS $p = 0.848$) and local differences near resonance is difficult to attribute to selection effects or sample composition, and instead points to a dynamical process that specifically reshapes the resonance architecture of valley-planet systems without displacing them to a different region of period ratio space. We discuss the physical implications in Section~\ref{sec:discussion}.
 
We show the period ratio distributions for both pair classes in Figure~\ref{fig:period_ratios}, with the locations of first-order mean-motion resonances indicated.

\subsection{Robustness tests}
\label{sec:robustness}

We conducted two tests to assess the robustness of our main finding that valley-inclusive pairs lack the peas-in-a-pod size-similarity enhancement observed in non-valley pairs.

\textbf{Valley membership criteria:} In Section~\ref{sec:relaxation}, we described the effects of relaxing our valley membership criteria on the qualitative radius ratio distributions. Table~\ref{tab:relaxation} quantifies the impact on our main statistical result. As valley criteria are relaxed, either by widening the radius window or by allowing larger fractional uncertainties, the sample size of valley-inclusive pairs increases from 38 to 89. However, the magnitude of the depletion weakens substantially, with $|\Delta|$ decreasing from 0.037 (default criteria) to 0.007 (widest criteria). This trend is consistent with dilution from planets whose valley membership is less secure. The signal is strongest among the most confidently identified valley planets (those satisfying the $\pm0.1\,R_\oplus$ window and $\leq5\%$ fractional uncertainty requirements), reinforcing our choice of conservative membership criteria.

\textbf{Pair separation:} To test whether the disruption of size-similarity is specific to physically adjacent planets, we repeated our analysis for non-adjacent pairs. For non-valley pairs, the peas-in-a-pod enhancement persists at non-adjacent separations ($n=256$, $\Delta = +0.069$, $p = 0.001$), though weaker than for adjacent pairs ($\Delta = +0.098$, $p < 0.001$). This is consistent with peas-in-a-pod being a system-level correlation that extends beyond immediate neighbors. For valley-inclusive pairs, non-adjacent pairs show a marginal positive enhancement ($n=85$, $\Delta = +0.052$, $p = 0.086$), in contrast to adjacent pairs which show no enhancement ($\Delta = -0.037$, $p = 0.858$). The reversal from depletion to marginal enhancement when moving from adjacent to non-adjacent separations indicates that the disruption of size-similarity is concentrated among the immediate neighbors of valley planets, as stated in Section~\ref{sec:relaxation}.

\begin{table}[h]
\begin{center}
\caption{Robustness of the valley-inclusive size-similarity result to alternative valley membership criteria}
\label{tab:relaxation}
\begin{tabular}{lcccccc}
\hline
Criterion & $n_\mathrm{valley}$ & $n_\mathrm{pairs}$ & $f_\mathrm{obs}$ & $f_\mathrm{null}$ & $\Delta$ & $p$ \\
\hline
Default ($\pm0.10\,R_\oplus$, $\leq5\%$) & 39 & 38 & 0.077 & 0.090 & $-0.037$ & 0.867 \\
$\pm0.15\,R_\oplus$ & 51 & 48 & 0.083 & 0.090 & $-0.007$ & 0.635 \\
$\pm0.20\,R_\oplus$ & 66 & 60 & 0.083 & 0.090 & $-0.007$ & 0.644 \\
$\leq7.5\%$ error & 96 & 89 & 0.067 & 0.089 & $-0.021$ & 0.813 \\
\hline
\end{tabular}
\end{center}
\end{table}


\section{Discussion}
\label{sec:discussion}
In this Discussion, we focus on the possibility that giant impacts play a non-trivial role in populating the radius valley, given the departure of radius valley planets from the parent distributions in radius and period ratio. First, we consider giant impacts as a driver for the shape of the radius ratio distribution in Section \ref{sec:structure}. In Section \ref{sec:dynamics}, we consider how giant impacts might contribute to the period ratio distribution among valley planets as well. We go on to speculate about the potential for supporting evidence among other dynamical properties.  
 
In considering the valley subsample, it is useful to first consider the radius and period ratios from the parent distribution. The peak of the distribution in $R_{\textrm{outer}}/R_{\textrm{inner}}$ at unity, with a tail towards increasing ratios indicates that the bulk of planetary systems show intra-system uniformity with a slight bias for planets growing larger at larger orbital distances. This is in line with expectations for planet populations largely sculpted by hydrodynamic atmospheric escape \citep{Owen13, Owen2017, Lopez2012, owen24}. We have identified moderate evidence that planets in the radius valley break this pattern. After accounting for the intrinsic rarity of valley planets, pairs involving a valley planet show no peas-in-a-pod enhancement above their shuffled planets null, in contrast to the strong enhancement among non-valley pairs. We consider the possibility that the normal mode of planet formation has been disrupted among valley planets. In particular, we argue that sculpting by giant impacts provides a plausible mechanism to reproduce both the radius ratio and period ratio findings. 
 
\subsection{Structure of radius ratio distribution}
\label{sec:structure}
 
We have shown that the peas-in-a-pod phenomenon appears to be inoperative among pairs involving a radius valley planet. non-valley pairs exhibit a $1.86\times$ enhancement of size-similar pairs above the baseline ($\Delta = +0.098$), while mixed pairs sit below the null ($\Delta = -0.037$). The widget test further excludes the possibility that the same mechanism operates at the same strength for mixed pairs (see Table~\ref{tab:widget_test}). This result accounts for the intrinsic rarity of valley planets by construction: the null for mixed pairs already reflects the scarcity of planets near $\sim 1.85\,R_\oplus$, so the absence of enhancement above this baseline is a statement about the physics of size-similarity, not about the shape of the radius distribution.
 
 Another way of expressing this finding is that radius valley planets tend to reside in one of two scenarios, distinct from self-similarity with neighbors. Typically either (1) the radius valley planet is the outer planet and exhibits reverse size-ordering, where it is smaller than its inner neighbor by $\sim30\%$, or else (2) the radius valley planet is an inner planet, but too small by 30\% to resemble its outer neighbor. 
 
In a scenario where radius valley planets do not exhibit self-similarity, the absence of peas-in-a-pod enhancement among valley planets suggests a departure. If planets in radius valley result from the same formation scheme, they would exhibit some degree of peas-in-a-pod enhancement above the null, as both super-Earths and sub-Neptunes do independently \citep{Millholland21, goyal_peas---pod_2024}. In contrast, a difference in radius of 30\% in either direction might indicate the presence of an unusually massive bare core: larger than typical super-Earths, but smaller than typical sub-Neptunes. Such a core otherwise ought to have retained an atmosphere under photoevaporation. \cite{Chance_2022} modeled how giant-impact driven mass loss versus photoevaporation ought to appear in radius/period space: they found that bare cores between 1.8--2.0$R_{\oplus}$, after acquiring a primordial atmosphere, are likely to retain it unless stripped by a giant impact. That is, bare rocky cores with radii of 1.8--2.0$R_{\oplus}$ are typically massive enough to keep their H/He atmospheres under photoevaporation. However, giant impacts are capable of stripping their primordial atmosphere. If distributed stochastically across planets, atmospheric loss via giant impacts could plausibly produce a lack of similarity with planets whose atmospheres are retained or lost according to the rules of photoevaporation.

Of course, if \textit{all} planets undergo giant impacts during an era of disruption, we would expect the stripped cores to resemble one another (see \citealt{izidoro_exoplanet_2022}). The radius ratio of unity is not entirely empty among valley planets: some mixed pairs do fall near $R_\mathrm{outer}/R_\mathrm{inner} = 1$
This is consistent with multiple pathways to the radius valley, with photoevaporation contributing some valley planets that retain correlated sizes and giant impacts contributing others that do not.
 
\subsection{Structure of period ratio distribution}
\label{sec:dynamics}
 
The period ratio distribution provides independent, dynamical evidence that radius valley planets have experienced a distinct evolutionary pathway. As shown in Section~\ref{sec:period_ratios}, the global period ratio distributions for mixed and non-valley pairs are statistically indistinguishable (KS $p = 0.848$). Mixed pairs cluster near the 3:2 resonance at more than twice the rate of non-valley pairs ($15.4\%$ versus $6.5\%$, $p = 0.09$), while showing a marginal deficit at tighter resonances ($2.6\%$ versus $10.5\%$, $p = 0.078$). Radius valley planets seem to avoid the tight orbital spacings that characterize a subset of the parent multi-transiting population (see Figure~\ref{fig:period_ratios}). For example, there are no pairs interior to the ``dynamically packed" stability criterion (at $ P_{\textrm{outer}}/P_{\textrm{inner}}\sim1.33$, per \citealt{wu_dynamical_2019}), compared to $5\%$ of non-valley pairs. One way of characterizing this pattern is that valley planets tend to lack very nearby neighbors unless the neighbor is in the 3:2 resonance.
 
This scarcity of close neighbors is paired with a slight over-representation of valley pairs near resonance.
To facilitate direct comparison with the results of \cite{dai_prevalence_2024}, we adopt their criterion for defining resonance proximity: planet pairs are considered near a MMR if their period ratios lie within the asymmetric boundary $-0.015 \leq \Delta \leq 0.03$ of exact period commensurability. The fraction of non-valley pairs near a first-order MMR is consistent with the rates reported by \cite{dai_prevalence_2024}. Among mixed pairs, the near-resonant fraction is elevated, driven almost entirely by clustering near the 3:2 resonance. Our findings partially contrast with those of \cite{dai_prevalence_2024}, who report a suppression of resonant structure among radius valley planets using a fixed $1.5$--$1.9\,R_\oplus$ cut. Differences in valley definitions, sample selection, and treatment of multiplicity likely contribute to this discrepancy.

While originally planets were anticipated to reside in long chains of mean motion resonances after disk dispersal (e.g. \citealt{terquem_migration_2007}), this is in stark contrast to the observations \citep{Lissauer11b, Fabrycky2014}. They may form in resonant chains, but they appear to drift from them over timescales of Gyr \citep{dai_prevalence_2024, hamer_kepler-discovered_2024, schmidt_resonant_2024}. However, quiescent migration to resonance is at odds with the observed radius ratios between valley planets that we describe here. Instead, peas-in-a-pod is typically \textit{stronger} among resonant pairs \citep{goyal_enhanced_2023}. In fact, previous studies in both exoplanetary radii and mass have demonstrated that peas-in-a-pod seems to persist regardless of proximity to MMR \citep{Millholland17, goyal_generalized_2022, wang_rv-detected_2017, otegi_similarity_2022}. In this sense, to see an overdensity of planets in resonance, but to have those planets exhibit unusual size-ordering, is difficult to explain from convergent type I migration alone.

A potentially more promising pathway toward a joint explanation of unusual size-ordering among valley planets, together with pile-ups in MMRs, is planet-planet scattering. A plausible path to a resonant system, determined from simulations by \cite{raymond_mean_2008}, involves close encounters between planets of disparate size. After a period of instability between one smaller and two larger planets, the smaller planet is usually ejected ($\sim $4 out of 5 times), but occasionally collides instead with one of the remaining planets. This process then leaves behind the pair of resonant planets. The scattering event may also imprint upon a difference in eccentricity among the surviving planets \citep{timpe_secular_2013}. A recent investigation of the objects in 3:2 resonance with Neptune by \cite{balaji_can_2023} indicated the higher likelihood of scattering inducing the resonant configuration, rather than smooth migration.

The contrast between global similarity and local divergence in the period ratio distribution strengthens this interpretation. If the anomalous resonance structure were driven by selection effects or by valley planets occupying a distinct region of period ratio space, we would expect the global distributions to differ. The KS test confirms that they do not ($p = 0.848$). The signal is instead localized to specific resonance neighborhoods, consistent with a dynamical process that reshapes the orbital architecture of valley-planet systems. The clustering near 3:2 in particular is consistent with a population that has been dynamically perturbed and subsequently relaxed toward the widest and most stable first-order resonance.
\begin{table}[ht]
\centering
\tablewidth{0pt}
\caption{Period ratio structure near first-order mean-motion resonances}
\label{tab:resonance_structure}
\begin{tabular}{cccc}
\hline
Resonance window & Valley-inclusive & Non-valley & $p$-value \\
\hline
Near 3:2                       & 6/39 (15.4\%)  & 50/769 (6.5\%)   & 0.09 \\
Near 5:4 + 4:3 + 5:3          & 1/39 (2.6\%)   & 81/769 (10.5\%)  & 0.078 \\
Below dynamical packing (1.33) & 0/39 (0\%)     & 39/769 (5.1\%)   & --- \\
\hline
\end{tabular}
\end{table}

\begin{figure*}[ht]
    \centering
    \includegraphics[width=0.9\textwidth]{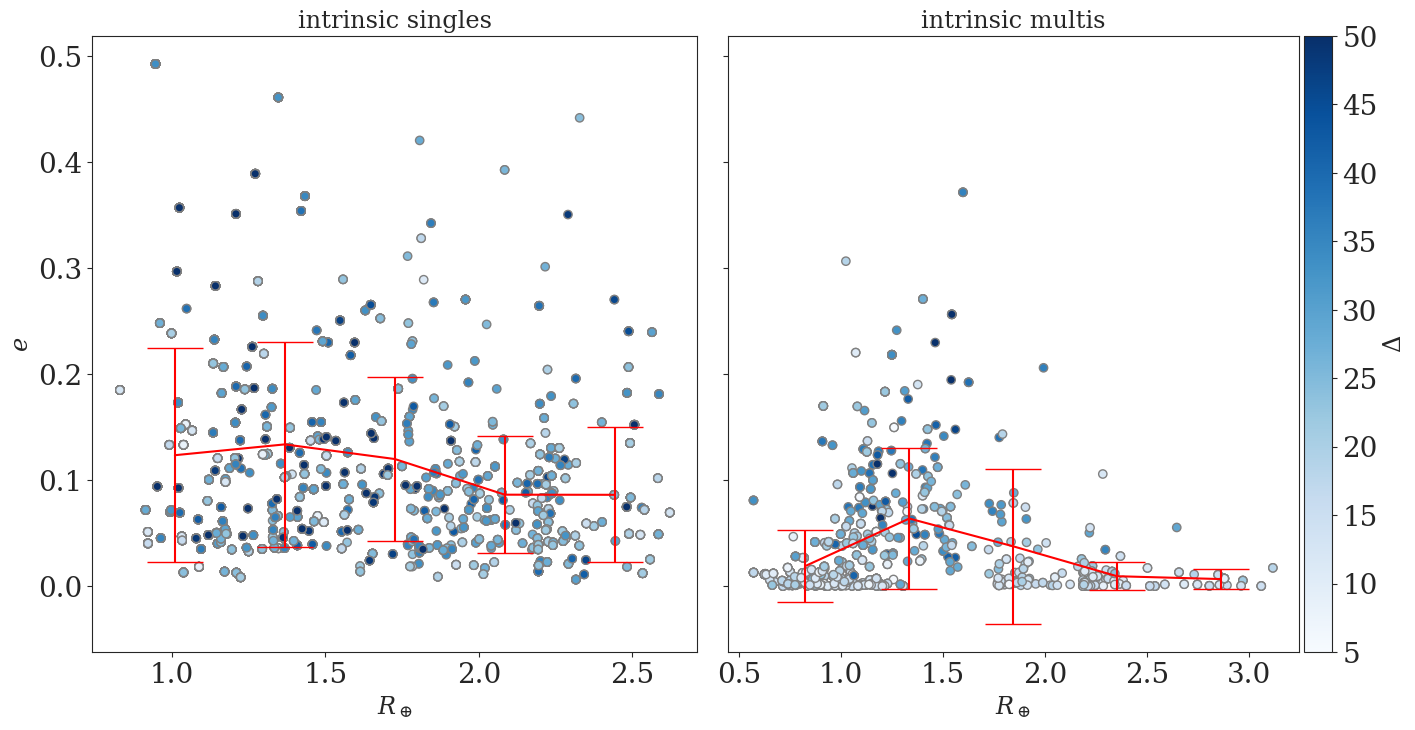}
    \includegraphics[width=0.9\textwidth]{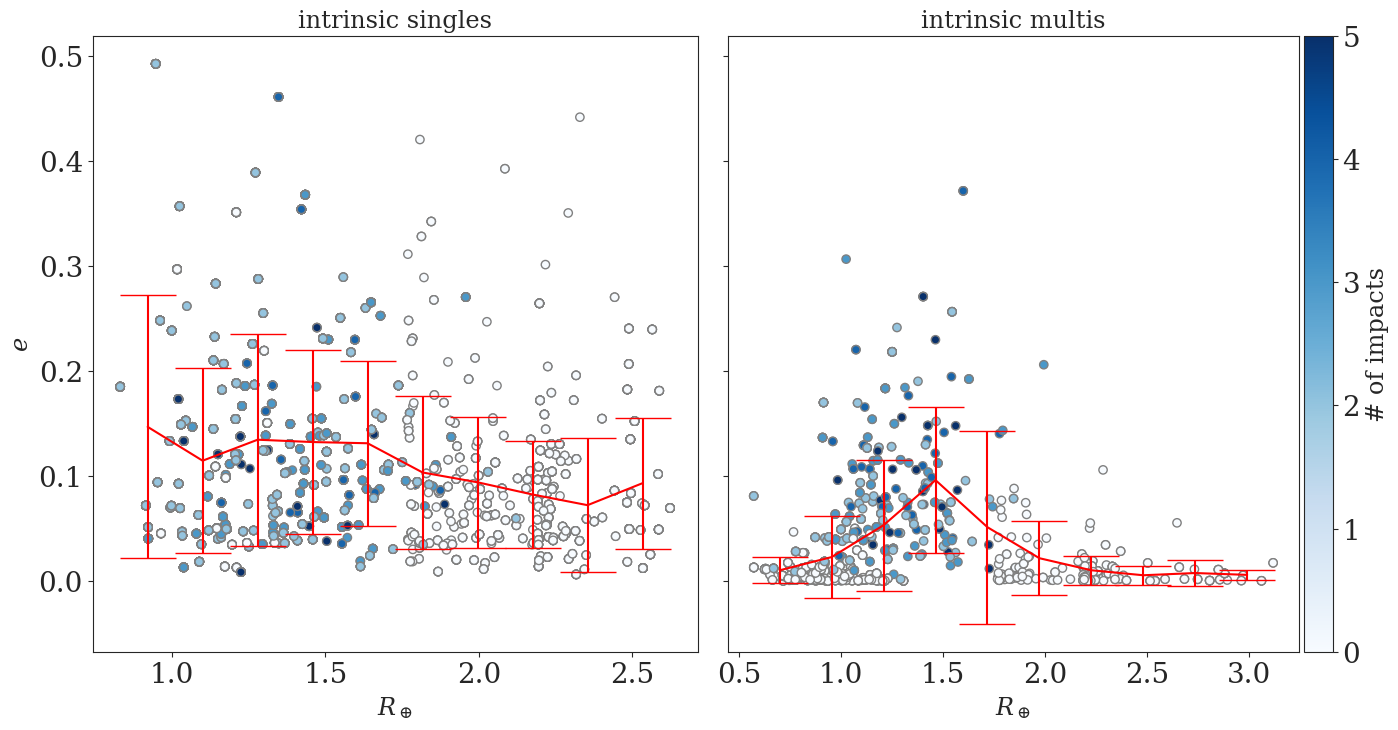}
 
    \caption{\textit{Top panels} Planetary radius versus eccentricity for planets from \cite{Chance_2022}, who employed the simulations of \cite{Dawson16} to model atmospheric loss from impacts. The simulations likeliest to produce the multi-transiting systems employed exclusively in this work are shown at left (colored by indicates mutual Hill spacing $\Delta$). The use of eccentricity on the y-axis is motivated  by the findings of \cite{gilbert_planets_2025}, which link increased eccentricity to planets in the radius valley. \textit{Bottom panel:} The same quantities, now color-coded to indicate the number of giant impacts after gas disk dispersal. The planetary systems resulting from the ``intrinsic multis" suite of simulations are generally dynamically cooler; except for planets in the radius valley; these are systematically more eccentric and more widely spaced from their neighboring planets.}
    \label{fig:simulations}
\end{figure*}

It is useful to turn to numerical simulations of planet formation to model expected observations; in this case, we revisit as a useful exercise the suite of simulations published by \cite{Dawson16}. \cite{Chance_2022} employed these simulations to model the effect of giant impacts upon the resulting radius distribution. That work contains a more detailed description of the simulations themselves, and the assumptions associated with mapping collision history to a radius distribution; here, we simply employ the same set of planets resulting from that work. In Figure~\ref{fig:simulations} we depict planetary radius versus orbital eccentricity for planets from two different simulation types. Intrinsic singles'' corresponds to the suite of simulations that tend to produce dynamically hotter systems: this is apparent from their overall higher eccentricity, as well as their typically higher mutual Hill spacing $\Delta$ (color-coded in legend). In contrast, intrinsic multis'' are drawn from simulations that produced dynamically colder outcomes: they tend to cluster around much lower eccentricities, and exhibit closer spacing. \cite{Dawson16} argued that this latter type of system is likelier the provenance of most of the multi-transiting systems observed by \textit{Kepler}, and correspond therefore to the planetary systems we consider in this work.

Planets near the radius valley (1.5-2$R_{\oplus}$) in these simulations have undergone multiple late-stage mergers with neighboring planets, leading to their anomalous radii and wider spacing from their neighbors. It is to be expected from this work that planets near the radius valley exhibit typically higher mutual Hill spacing, with eccentricity increasing dependent upon the number of giant impacts experienced by the planet. We see some supporting evidence for increased mutual Hill spacing among planets that experienced a giant impact: they tend to possess average mutual Hill spacing closer to 30, as compared to the typical 20 (consistent with \citealt{Weiss2018}).

From a cursory consideration of \cite{Chance_2022}, which focused specifically on the demographics of radius valley planets, we can predict that they ought to possess both high mutual Hill spacing and eccentricity. With respect to the former, we argue there is some observable supporting evidence for this phenomenon, given the dearth of close-in neighbors to valley planets \textit{not} in an MMR found in this study and the increase in gap complexity for radius valley planets found in \cite{rice_distribution_2024}. With respect to the latter, recent work by \citet{gilbert_planets_2025} provides supporting evidence: they find that planets residing in the radius valley exhibit systematically higher eccentricities than their neighbors, consistent with expectations from late-stage giant impacts.

If indeed eccentricities for planets in the radius valley are higher, the pile-up in resonance is perhaps consistent: such resonances might either (1) provide the gravitational perturbations needed to maintain their eccentricity over long timescales (see e.g. \citealt{peale_orbital_1976}) or simply (2) be the only surviving configurations if the eccentricity is high enough to threaten orbit-crossing. Generally speaking, therefore, the more eccentric the planets of a system are, the more likely they will be in resonance \citep{bailey_period_2022}. 
 
\section{Conclusions}
\label{sec:conclusions}

Using the large sample of \textit{Kepler} planets with precise radii from the \cite{berger_gaia-kepler-tess-host_2023} catalog, we have investigated the peas-in-a-pod phenomenon for the population of planets residing in the radius valley. The population likely represents the intersection of many different planet formation and evolution processes, and valley planets in particular can be diagnostic of formation mechanisms. We identify, using a difference-in-differences test that accounts for the intrinsic rarity of valley planets, that the peas-in-a-pod size-similarity enhancement is entirely absent among pairs involving a radius valley planet, while non-valley pairs exhibit strong enhancement. The difference between the two pair classes is significant 
(p = 0.0002), and we can positively exclude that the same enhancement operates for 
valley-inclusive pairs at the strength observed in the parent population ($p < 0.001$). Rather, radius valley planets are likelier to reside in configurations where they differ from their neighbors by $\sim 30\%$ in radius, with the distribution of $R_\mathrm{outer}/R_\mathrm{inner}$ peaking between $0.7$ and $0.8$, and a secondary peak at $\sim 1.3$.
 We caution that our valley sample is modest in size ($n = 39$ pairs), and future studies with larger samples will be essential to confirm the robustness of the disrupted size-similarity signal.
We also identify that the period ratio distributions for mixed and non-valley pairs, while globally indistinguishable (KS $p = 0.848$), exhibit strikingly different local structure. Mixed pairs cluster near the 3:2 resonance at more than twice the rate of non-valley pairs ($15.4\%$ versus $6.5\%$), while showing a marginal deficit at tighter resonances and a complete absence of pairs below the dynamical packing limit of $P_\mathrm{outer}/P_\mathrm{inner} = 1.33$.
 
Taken together, the distinct architectural signatures of radius valley planets, including suppressed self-similarity, enhanced 3:2 resonance clustering, and elevated eccentricities \citep{gilbert_planets_2025}, suggest that their formation or evolution has differed meaningfully from that of the broader planet population. No single piece of evidence is conclusive on its own, but the convergence of three independent lines of evidence from radius ratios, period ratios, and eccentricities is difficult to reconcile with photoevaporation or core-powered mass loss as the sole origin of valley planets. These mechanisms act deterministically as a function of planet mass, orbital period, and stellar irradiation, and do not predict disrupted size-ordering, anomalous resonance architecture, or elevated eccentricities. Instead, the observations point toward a stochastic process, such as late-stage giant impacts, capable of simultaneously stripping atmospheres and altering system architectures. The apparent dynamical peculiarity of these systems raises the possibility that planets in the radius valley may not represent only the overlap of the terrestrial and sub-Neptune populations, but also a distinct evolutionary pathway. As a relatively rare and structurally unusual population, planets in the radius valley may serve as key pieces of evidence in understanding the diversity of planet formation outcomes.

\section*{Acknowledgements}
 
Q.C.'s work was supported by the National Aeronautics and Space Administration under Grant No. 80NSSC21K1841 issued through the Future Investigators in NASA Earth and Space Science and Technology program.
Q.C. acknowledges support from the NASA Postdoctoral Program at NASA Goddard Space Flight Center, administered by Oak Ridge Associated Universities under contract with NASA.
 
This research has made use of the NASA Exoplanet Archive, which is operated by the California Institute of Technology, under contract with the National Aeronautics and Space Administration under the Exoplanet Exploration Program.
 
We are thankful for helpful discussions and input from Hilke Schlichting, Gregory Gilbert, and Chris Lam. 
 
\paragraph{Software}
    astropy \citep{astropy_collaboration_astropy_2013,astropy_collaboration_astropy_2018}, numpy \citep{harris_array_2020}, matplotlib \citep{Hunter:2007}, pandas \citep{reback2020pandas, mckinney-proc-scipy-2010},
          ChatGPT (OpenAI, 2025),
          Gemini (Google DeepMind, 2025)

\label{sec:Acknowledgements}
 
\bibliography{merged_clean_references}

\end{document}